\documentclass[aps,prb,reprint,groupedaddress,showpacs]{revtex4-1}

\usepackage[utf8]{inputenc}
\usepackage{color}
\usepackage{soul}
\usepackage{dcolumn}
\usepackage{multirow}
\usepackage{amsmath}
\usepackage{graphicx}
\usepackage[unicode=true,pdfusetitle,bookmarks=true,bookmarksnumbered=false,bookmarksopen=false,breaklinks=true,pdfborder={0 0 1},backref=false,colorlinks=true]{hyperref}
\hypersetup{allcolors=blue}

\begin{document}


\title{Strain-induced structure transformations on Si(111) and Ge(111) surfaces: a combined density-functional and scannning tunnneling microscopy report}


\author{R. Zhachuk}\email{zhachuk@gmail.com}
\affiliation{Institute of Semiconductor Physics, pr. Lavrentyeva 13, Novosibirsk 630090, Russia}

\author{S. Teys}
\affiliation{Institute of Semiconductor Physics, pr. Lavrentyeva 13, Novosibirsk 630090, Russia}

\author{J. Coutinho}
\affiliation{I3N, Department of Physics, University of Aveiro, Campus Santiago,
3810-193 Aveiro, Portugal}

\date{\today}

\begin{abstract}
Si(111) and Ge(111) surface formation energies were calculated using density functional theory for various biaxial strain states ranging from $-0.04$ to $0.04$, and for a wide set of experimentally observed surface reconstructions: $3\times3$, $5\times5$, $7\times7$ dimer-adatom-stacking fault reconstructions and $c(2\times8)$, $2\times2$ and $\sqrt{3}\times\sqrt{3}$ adatoms based surfaces. The calculations are compared with scanning tunneling microscopy data obtained on stepped Si(111) surfaces and on Ge islands grown on a Si(111) substrate. It is shown that the surface structure transformations observed in these strained systems are accounted for by a phase diagram that relates the equilibrium surface structure to the applied strain. The calculated formation energy of the unstrained Si(111)-$9\times9$ dimer-adatom-stacking fault surface is reported for the first time and it is higher than corresponding energies of Si(111)-$5\times5$ and Si(111)-$7\times7$ dimer-adatom-stacking fault surfaces as expected. We predict that the Si(111) surface should adopt a $c(2\times8)$ reconstruction when tensile strain is above $0.03$.
\end{abstract}


\maketitle

\section{Introduction}

During heteroepitaxial growth, elastic strain appears when the substrate and growing film lattice constants mismatch. This is evident during Ge/Si growth and in the vicinity of surface defects like step edges and other sites where preferential relaxation occurs. Strain also has great impact on surface structure, diffusion of adsorbed atoms, thin film growth, nanostructure formation and stabilization of facet planes.\cite{ibach-ssr-29-193,*cammarata-pss-46-1,*muller-pss-54-157,*cherepanov-apl-81-4745,*persichetti-prb-85-195314,*raiteri-prl-88-256103,*fujikawa-prl-88-176101}

A typical example of a strained system is germanium on silicon. Interest in Ge heteroepitaxial growth on Si stems from several reasons. First, it has shown great potential for new high frequency semiconducting devices, while keeping compatibility with the mature Si-based complementary metal-oxide-semiconductor technology. Second, the recent success in growing structurally perfect Ge/Si nanowires and nanorings, may provide the structural building blocks for future nanoelectronic devices.\cite{kawamura-prl-91-096102} Third, it is a prototypical system for studying strained Stranski-Krastanov (SK) growth. Accordingly, after the first few smoothly grown epitaxial layers (wetting layer), growth proceeds with the formation of islands (SK growth mode).\cite{bechstedt-psp-2003} This is accompanied by several reconstruction changes on top of the growing Ge islands.\cite{motta-ss-406-254} However, technological difficulties exist due to the fact that Si and Ge lattice constants differ by about 4\%. This leads to a build-up compressive stress on Ge, making the growth of defect-free Ge films on Si a rather difficult task.

Although it is commonly accepted that an applied elastic strain can induce surface structure transformations, precise data quantifying this effect are yet unavailable.\cite{gossmann-prl-55-1106} The reported calculated data for Si(111) and Ge(111) are either based on simplified surface reconstruction models (which did not include adatoms and other important structural features) or they were obtained using empirical potential and tight binding calculations.\cite{vanderbilt-prl-59-1456,mercer-prb-48-5374} These methods not only are limited with respect to their prediction accuracy of surface absolute formation energies, but also can hardly account for the strain response of complex surface reconstructions and hybridizations. Hence, the need for further progress in this field, in particular the lack of surface-strain phase diagrams for Si(111) and Ge(111), calls for a study of these effects by means of state-of-the-art first principles calculations.

Having said that, this work aimed at establishing a correspondence between Ge(111) and Si(111) surface structure changes and the applied lateral strain by combining density functional theoretical (DFT) calculations with scanning tunneling microscopy (STM) data. Si and Ge surfaces were calculated independently. With this approach we overlooked any species intermixing that may occur at the interface during growth. Nevertheless, given the chemical resemblance between both species, this effect should have a limited impact.

The article is organized in the following manner -- in Section~\ref{sec:methods} we describe the computational details, convergence tests to the calculations, as well as the experimental procedure to obtain the reported STM data; Section~\ref{sec:results} reports the calculations of surface energies and surface strain response, as well as a discussion of these results with the STM data as scenery; and finally we summarize the conclusions in Section~\ref{sec:conclusions}.

\section{Methods}\label{sec:methods}

\subsection{Computational details}

Absolute surface formation energies were calculated according to the recipe by Stekolnokov \emph{et al.}\cite{stekolnikov-prb-65-115318}, here adapted for strained surfaces. Accordingly, the surface formation energy of some reference surface should be calculated using a slab with two equivalent surfaces (symmetric slab). Relaxed unreconstructed $1\times1$ surfaces of Si(111) and Ge(111) were chosen for such references. The layer in the middle of the slab was kept frozen, while atoms in other layers were allowed to move without any constraints during atomic optimizations. The surface formation energy per unit area ($\gamma$) is obtained from the energy excess of the slab compared to that of a bulk calculation with same number of atoms. Hence, for reference surfaces

\begin{equation}
\gamma_{1\times1}(\epsilon)=\frac{1}{2 S_{1\times1}(\epsilon)}
\left[ E_\mathrm{tot}^{1\times1}(\epsilon)-\mu(\epsilon) N \right],
\end{equation}

where $\mu$ is the Si or Ge strain-dependent chemical potential, \emph{i.e.} the energy per atom in bulk under strain $\epsilon$, $S_{1\times1}$  is the area of a $1\times1$ surface unit cell and $E_\mathrm{tot}^{1\times1}$ is the total energy of the slab comprising $N$ atoms per simulation cell. The factor of $1/2$ avoids double counting both slab surface energies.

According to Ref.~\onlinecite{stekolnikov-prb-65-115318}, the calculation of the energy gain due to surface reconstructions requires using hydrogenated slabs, where one surface layer (hereafter referred as bottom layer) is saturated with hydrogen atoms, while the reconstructions are considered at the opposite surface. Two slabs should be used in this approach -- one for the chosen reference (in our case the unreconstructed $1\times1$ surface), and a second slab for the reconstructed surface under scrutiny. With this setup the location of H and Si (Ge) atoms at the bottom layer is kept frozen during atomic optimizations, while all other atoms are freely allowed to relax. Thus, the energy gain per unit area is now

\begin{equation}
\Delta\gamma_\mathrm{rec}(\epsilon)=\frac{1}{S_\mathrm{rec}(\epsilon)}
\left[ E_\mathrm{tot}^\mathrm{rec}(\epsilon)-
       E_\mathrm{tot}^{1\times1}(\epsilon) M - \mu(\epsilon) K \right],
\end{equation}
where $S_\mathrm{rec}$ is the unit cell area of the reconstructed slab, $M=S_\mathrm{rec}/S_{1\times1}$ is the number of $1\times1$ reference cells per reconstructed unit cell, and $K$ accounts for the number of Si (Ge) surface atoms in excess to those in the reference cell. The absolute surface formation energy of a reconstructed surface under strain is obtained after combining the energy reference from the unreconstructed $1\times1$ surface with the energy gain due to reconstructions:

\begin{equation}
\gamma_\mathrm{rec}(\epsilon)=\gamma_{1\times1}(\epsilon)+\Delta\gamma_\mathrm{rec}(\epsilon).
\end{equation}

Total energies were calculated from first principles by using the local density functional \textsc{siesta} code.\cite{soler-jpcm-14-2745} The exchange and correlation functional is that of Ceperley-Alder as parametrized by Perdew and Zunger.\cite{perdew-prb-23-5048} The $\mathbf{k}$-space integrals over three- or two-dimensional Brillouin zones (BZ’s) were approximated by sums over Monkhorst-Pack (MP) type special points.\cite{monkhorst-prb-13-5188} Norm-conserving pseudopotentials were employed to account for electronic core states,\cite{troullier-prb-43-1993} whereas valence states are spanned with help of linear combinations of numerical atomic orbitals of the Sankey-Niklewski type, generalized to be arbitrarily complete with the inclusion of multiple $\zeta$ orbitals and polarization states.\cite{soler-jpcm-14-2745} For convergence control purposes, large basis calculations were carried out by assigning polarized double-$\zeta$ functions (DZP) to all atoms in the slab. This means two sets of $s$ and $p$ orbitals plus one set of $d$ orbitals on Si and Ge atoms, and two sets of $s$ orbitals plus a set of $p$ orbitals on H. To make the problem computationally treatable, most calculations were performed using single-$\zeta$ functions (SZ) for H as well as Si and Ge atoms at the three bottom slab layers, while DZP basis was kept on atoms at the remaining layers. Si and Ge atoms with SZ basis have one set of $s$ and $p$ orbitals, while H atoms have a single $s$ orbital. Results obtained by using such a basis combination are labeled with SZ-DZP.

%
\begin{figure}
\includegraphics[width=7.5cm]{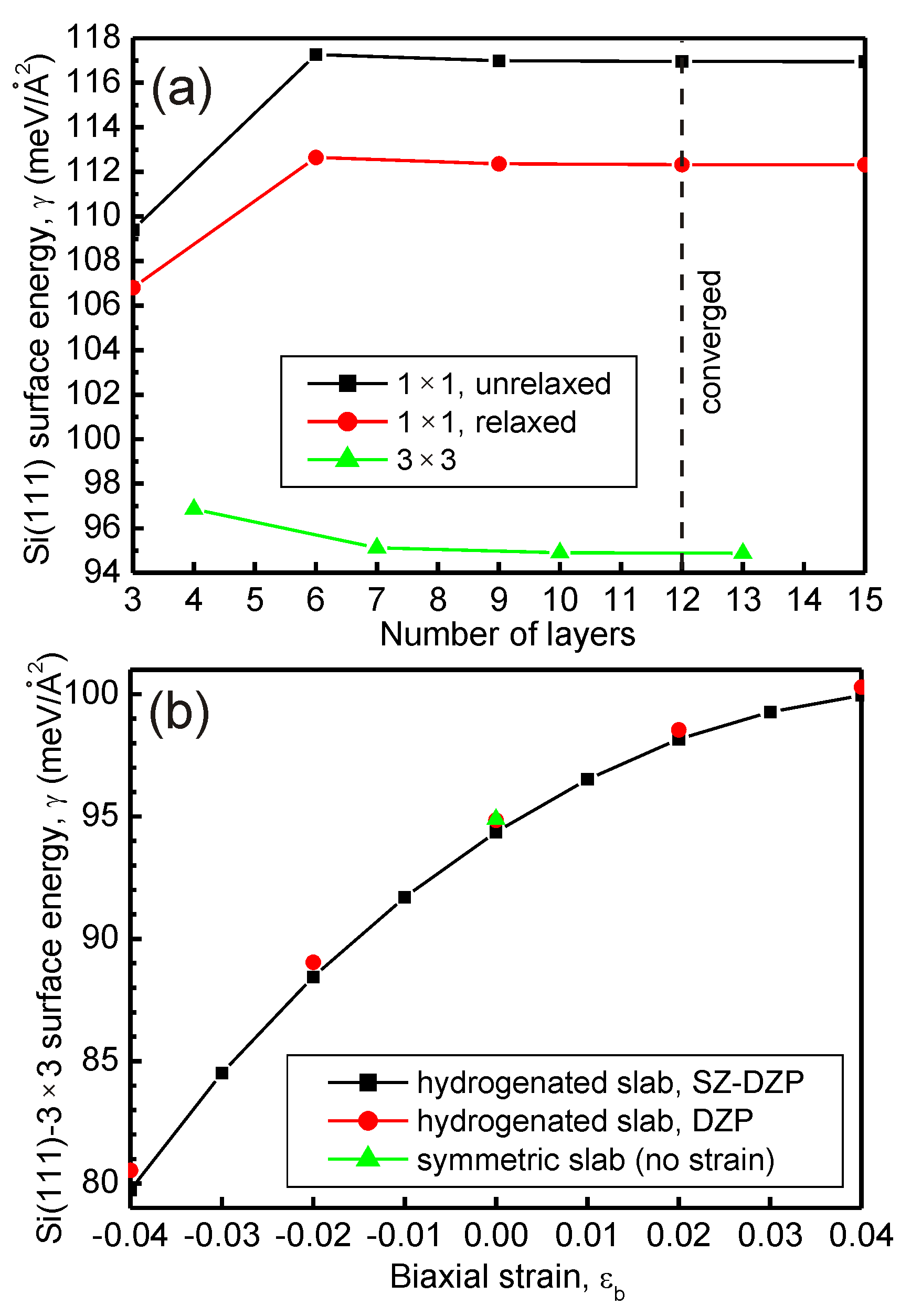}
\caption{\label{fig1}(Color online) (a) Surface formation energy as a function of the number of Si(111) layers in symmetric slabs with $1\times1$ and $3\times3$ DAS surface reconstructions. (b) Si(111)-$3\times3$ DAS surface formation energy as a function of applied biaxial strain $\epsilon_\mathrm{b}$, calculated using a 6-layer thick hydrogenated slabs with DZP and mixed SZ-DZP basis sets.}
\end{figure}

The unreconstructed bottom surfaces were saturated by hydrogen atoms making 1.50~{\AA} Si-H and 1.55~{\AA} Ge-H bonds. The opposite slab surfaces were set up according to specific surface structure models. These are dimer-adatom-stacking fault (DAS) based models, as well as adatoms-based $c(2\times8)$, $2\times2$ and $\sqrt{3}\times\sqrt{3}$ reconstructions. The DAS models were initially proposed by Takayanagi \emph{et al.}\cite{takayanagi-ss-164-367} for Si(111)-$7\times7$ and later extended for other DAS-family member structures ($3\times3$, $5\times5$, $7\times7$, $9\times9$, etc.).\cite{becker-prb-39-1633} Adatoms on $c(2\times8)$, $2\times2$, and $\sqrt{3}\times\sqrt{3}$ surfaces were placed at high-symmetry $T_4$ adsorption sites. This is well established for adatoms on the Ge(111)-$c(2\times8)$ reconstruction.\cite{becker-prb-39-1633} We also know that $T_4$ adsorption sites are energetically more favorable than $H_3$ and $T_1$ in other cases.\cite{qin-prb-75-085313} The $c(2\times4)$ reconstruction, observed on quenched Si(111) surfaces and on Ge/Si(111) during growth was not considered. Our experimental results indicate that it occupies a small fraction area between different $2\times2$ or $c(2\times8)$ domain structures and therefore is considered a domain wall structure.\cite{kohler-ss-248-321}

A uniform real-space grid equivalent to a plane-wave cutoff of 200~Ry was used for Fourier transforming the density and potential terms. The geometry was optimized until all atomic forces became less than 0.01~eV/{\AA}, when the surface structure was considered to have attained equilibrium. All periodic slabs were separated by a 30~{\AA} thick vacuum layer. Under these conditions, converged calculations using bulk unit cells with a MP-$8^3$ BZ-sampling scheme gave lattice constants of Si and Ge were $a_\mathrm{Si}=5.420$~{\AA} and $a_\mathrm{Ge}=5.650$~{\AA}. These are in reasonable agreement with the experimental values $a_\mathrm{Si}=5.430$~{\AA} and $a_\mathrm{Ge}=5.660$~{\AA}, respectively.

\subsection{Convergence tests}

Here we describe the convergence tests to the conditions that have most impact on absolute surface formation energies, starting with BZ sampling. To determine the optimum BZ sampling scheme for a particular slab lattice, we found the converged value of the Si (Ge) chemical potential $\mu$ from a bulk supercell employing a lattice identical to that of the slab (these geometries are hereafter termed as \emph{bulk-slabs}). To this end we progressively increased the density of the $\mathbf{k}$-point grid until $\mu$ changed by less than 5~$\mu$eV/atom. Such a grid was considered to be appropriate for slab calculations with surface reconstructions (and vacuum gap).

For Si and Ge bulk unit cells at zero strain, $\mu$ converged at $-107.844476$~eV and $-108.576450$~eV, respectively, using a MP-$8^3$ grid of $\mathbf{k}$-points. However, in order to reduce errors in relative surface energies, all calculations used a common value of $\mu$ obtained from a unitary bulk-slab, 3 layers thick and $1\times1$ in the (111) plane. With this geometry and MP-$20\times20\times8$, Si and Ge chemical potentials converged at $\mu=-107.844483$~eV and $\mu=-108.576456$~eV, respectively, differing from the unit cell values by less than 7~$\mu$eV/atom only. After carefully looking at the convergence of $\mu$ using bulk-like slabs, we ended up with the following BZ sampling schemes for reconstructed surface slabs: MP-$20\times20\times1$ for $1\times1$, MP-$6\times6\times1$ for $3\times3$ DAS, MP-$4\times4\times1$ for $5\times5$ DAS, MP-$3\times3\times1$ for $7\times7$ DAS, MP-$2\times2\times1$ for $9\times9$ DAS, MP-$10\times10\times1$ for $2\times2$, MP-$8\times2\times1$ for $c(2\times8)$ (rectangular surface cell) and MP-$12\times12\times1$ for $\sqrt{3}\times\sqrt{3}$. The resulting $\mathbf{k}$-point densities in reciprocal space are approximately the same for all schemes. The estimated error in surface formation energies when compared to calculations with denser grids is well below 0.1~meV/{\AA}$^2$ for all reconstructions.

\begin{figure}
\includegraphics[width=8.8cm]{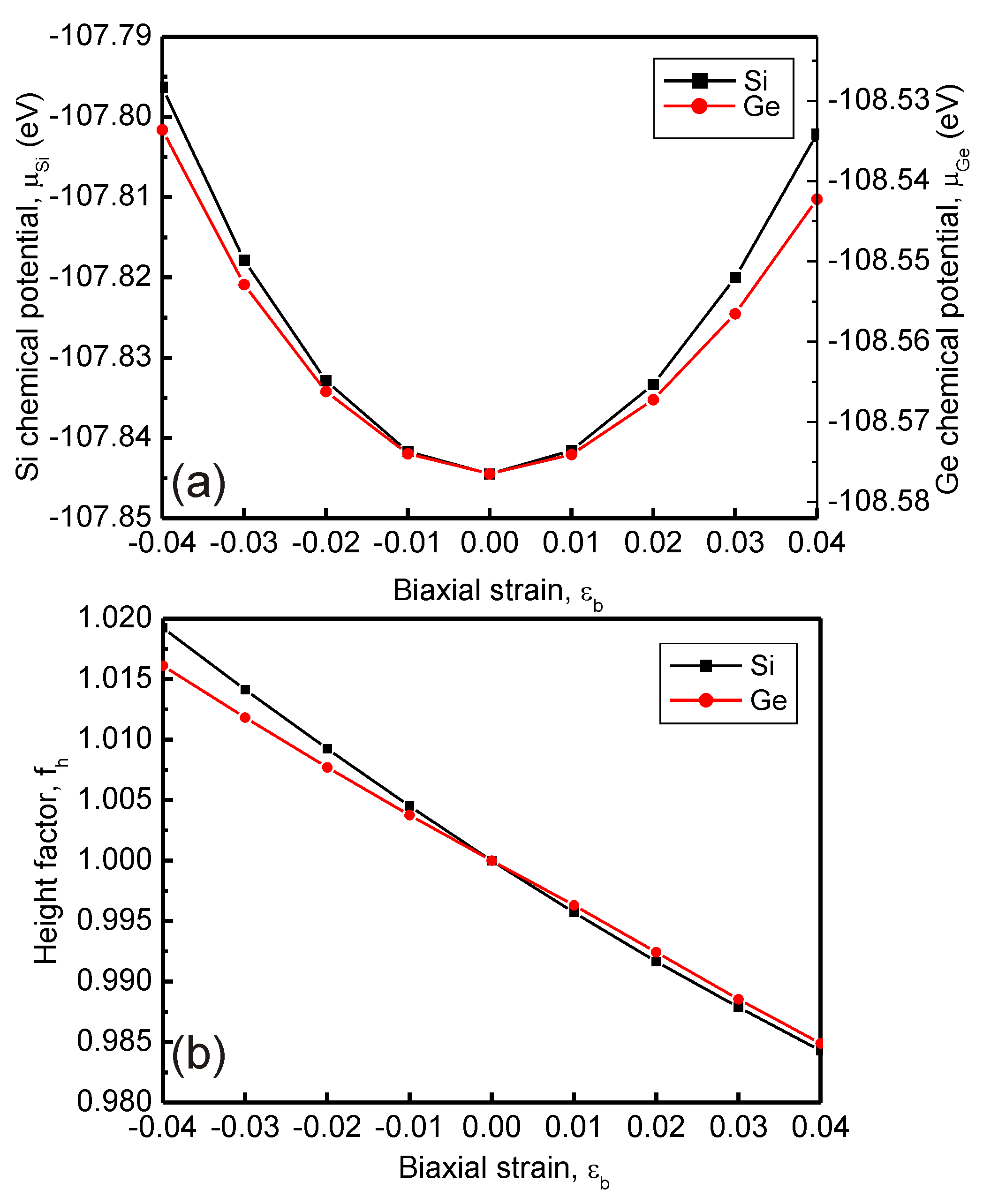}
\caption{\label{fig2}(Color online) (a) Si and Ge chemical potentials and (b) height factors, $f_\mathrm{h}$ (see text) for Si(111) and Ge(111) interplanar spacings, calculated for various biaxial strain states $\epsilon_\mathrm{b}$ in the (111) plane.}
\end{figure}

Figure~\ref{fig1}(a) shows the calculated surface energy using DZP basis functions, for the ideal Si(111)-$1\times1$ (relaxed and unrelaxed) and reconstructed Si(111)-$3\times3$ DAS surfaces as a function of the number of atomic layers in symmetric slabs. Beyond 12 layers the surface energy is converged within 0.01~meV/{\AA}$^2$. These were therefore the number of layers used to calculate the reference surface energy $\gamma_{1\times1}(\epsilon)$ from a Si(111)-$1\times1$ symmetric slab. Figure~\ref{fig1}(b) depicts the Si(111)-$3\times3$ DAS surface energy calculated using hydrogenated slabs with full DZP and mixed SZ-DZP basis as a function of biaxial surface strain. The replacement of DZP by mixed SZ-DZP basis causes a considerable decrease in the size of the Hamiltonian, at the expense of a systematic downward shift of surface energies by about 0.5 meV/{\AA}$^2$ across all strain window of interest. Accordingly, we used DZP basis functions on Si (Ge) atoms belonging to the 3 upper layers, while cheaper SZ basis sets were assigned both to atoms at the lowest 3 layers as well as to the hydrogen species. Although this step introduces the largest error in the calculated surface formation energies -- about 0.5~meV/{\AA}$^2$, it has little impact to the calculations of the surface strain response.

Since most calculations were performed using hydrogenated slabs, we estimated the error produced by this approach. Usage of hydrogenated slabs requires less atomic layers, leading to a considerable reduction in computational effort in achieving self-consistent electron densities and energies.\cite{stekolnikov-prb-65-115318} We found that for hydrogenated slabs, 6 Si (Ge) layers were sufficient to provide highly accurate surface formation energies. The surface energy difference between 6-layered hydrogenated and 12-layered symmetric Si(111)-$3\times3$ DAS surfaces is only 0.03~meV/{\AA}$^2$. Hence, the total number of atoms in hydrogenated slabs were the following: 13 for $1\times1$, 133 for $3\times3$ DAS, 325 for $5\times5$ DAS, 641 for $7\times7$ DAS, 1063 for $9\times9$ DAS, 53 for $2\times2$, 212 for $c(2\times8)$ and 40 for $\sqrt{3}\times\sqrt{3}$. For symmetric slabs the number of atoms was 30 for $1\times1$ and 248 for $3\times3$ DAS. The surface energy for the $9\times9$ DAS reconstruction was calculated for zero strain only. A strain analysis for such a slab would require an incommensurate computational effort.

\subsection{Experimental procedure}

The experiments were performed in a ultra-high vacuum system equipped with a STM (\textsc{omicron}). The $12\times3\times0.4$~mm$^3$ silicon samples were cut out from n-type silicon wafers with a resistivity of 0.5~$\Omega\,$cm. The samples were heated by current injection and their temperature was controlled with help of an optical disappearing filament pyrometer.

The surface cleaning procedure was carried out at a pressure of $2\times10^{-10}$~Torr. The Si(111) samples were cleaned by degassing at least for 4~h at 600$^\circ$C followed by flash annealing at 1250$^\circ$C for a few seconds. We cooled the samples stepwise with 50$^\circ$C per minute steps within the temperature range 900–350$^\circ$C by reducing the heating current. Clean Si($7\,7\,10$) surfaces were prepared following the procedure described by Kirakosian \emph{et al.}\cite{kirakosian-apl-79-1608} Firstly, a sample was carefully degassed at 600$^\circ$C for several hours. Then, after short flash annealing at 1250$^\circ$C, the temperature of the sample was lowered to 1060$^\circ$C during 30~s. After that the sample was quenched to 830$^\circ$C, kept at this temperature for 15~min, and finally cooled down to room temperature during 20~min.

The source of germanium atoms consisted of pieces of Ge fastened to a W ribbon. The ribbon was heated upon current injection. Hereafter we specify the Ge deposited coverage and thickness of Ge islands in bilayer units (1~{BL~Ge}$=1.44\times10^{15}$~at/cm$^2$). These are equivalent to full Si (Ge) layers and must be distinguished from monolayers (ML), which are $1~\mathrm{ML}=1/2~\mathrm{BL}$ for (111) surfaces of crystals with the diamond structure. This has been commonly adopted among experimentalists working with Ge/Si(111). The Ge deposition rate in our experiments was 0.001-0.01~BL/min. The Ge flux was calibrated by measuring the total volume of Ge islands formed on Si(111) far from step edges and domain boundaries of surface reconstructions.\cite{teys-pls-1-37}

STM images were recorded at room temperature in the constant-current mode using an electrochemically etched tungsten tip. The freely available \textsc{wsxm} software was used to analyze and process the STM images.\cite{horcas-rsi-78-013705}

\section{Results and discussion}\label{sec:results}

\subsection{Surface energies and stresses of Si(111) and Ge(111)}

\begin{figure}
\includegraphics[width=7.3cm]{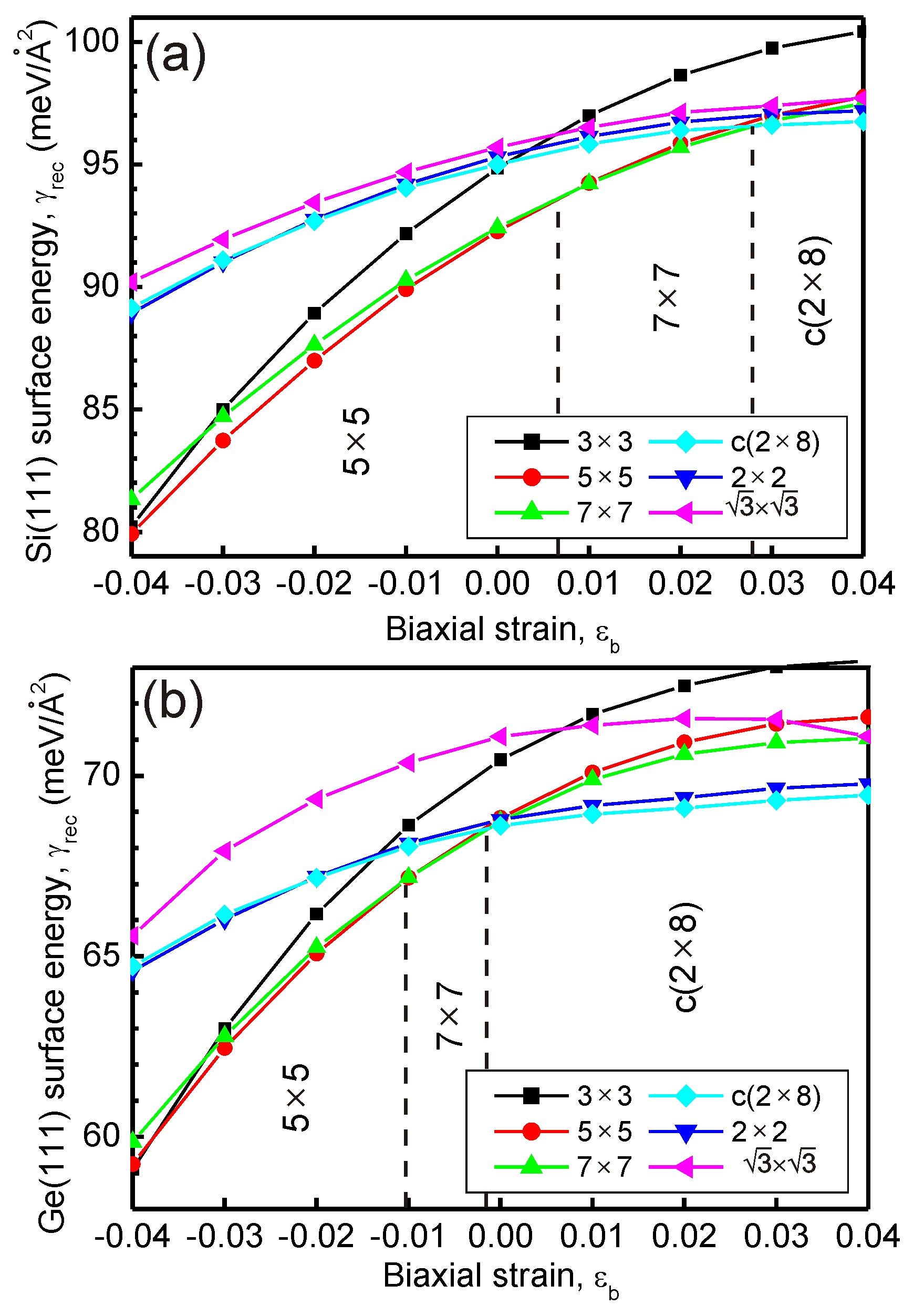}
\caption{\label{fig3}(Color online) Surface formation energies calculated for various biaxial strain states $\epsilon_\mathrm{b}$ and surface reconstructions for (a) Si(111) and (b) Ge(111) surfaces.}
\end{figure}

With the exception of the $c(2\times8)$ surface structure, for which a small anisotropy on its surface stress tensor is expected, all other surface reconstructions in this study have $C_3$ symmetry. For such surfaces the surface stress tensor is isotropic: $\sigma_{xx}=\sigma_{yy}=\sigma_\mathrm{rec}$ and $\sigma_{xy}=\sigma_{yx}=0$, where the $[111]$ direction is assumed to be parallel to the $z$-axis. Hence, we applied an isotropic biaxial strain $\epsilon_{xx}=\epsilon_{yy}=\epsilon_\mathrm{b}$, and from the surface formation energy $\gamma_\mathrm{rec}(\epsilon_\mathrm{b})$ we found the surface stress $\sigma_\mathrm{rec}(\epsilon_\mathrm{b}=0)$. For surfaces having $C_3$ symmetry the dependence of $\gamma_\mathrm{rec}$ on uniaxial strain $\epsilon_\mathrm{u}$ can be achieved from $\gamma_\mathrm{rec}(\epsilon_\mathrm{b})$ by using the relation,

\begin{equation}\label{eq:bi-uni}
1+\epsilon_\mathrm{u}=\left( 1+\epsilon_\mathrm{b} \right)^2.
\end{equation}

Figure~\ref{fig2}(a) shows the calculated chemical potential $\mu$ as a function of (111) biaxial strain, $\epsilon_\mathrm{b}$, for bulk Si and Ge. From elasticity it follows that a (111)-biaxially strained cubic solid suffers an opposite strain along $[111]$. Such effect has to be accounted for in strained surface calculations, and so one should let the surfaces relax towards the vacuum. This effect was considered by using appropriately strained $1\times1$ bulk-slabs with several heights, $h$, related to the strain-free height $h_0$ by a \emph{height factor}, $f_\mathrm{h}=h/h_0$. We determined their equilibrium heights ($h$ values that minimized the energy), which were then used to obtain the chemical potentials of Si (Ge) for a specific strain state. Fig.~\ref{fig2}(b) depicts calculated height factors for Si and Ge $1\times1$ bulk-slabs. It shows how the equilibrium distance between $(111)$ layers in bulk depends on the applied biaxial strain. A direct comparison between Si- and Ge-related plots clearly shows that in agreement with the experimental data, Si has greater mechanical stiffness than Ge. Both Fig.~\ref{fig2}(a) and Fig.~\ref{fig2}(b) show that the plots are slightly asymmetric and hence the strain interval spans the anharmonic regime.

Figures~\ref{fig3}(a) and \ref{fig3}(b) show the calculated surface energies $\gamma_\mathrm{rec}(\epsilon_\mathrm{b})$ of Si(111) and Ge(111), respectively, for various experimentally observed reconstructions. These were calculated using combined SZ-DZP basis for $\epsilon_\mathrm{b}$ values between $-0.04$ to 0.04, and shifted by their respective systematic errors with respect to the more complete DZP basis calculations at $\epsilon_\mathrm{b}=0$. Calculations of the unreconstructed Si(111)-$1\times1$ are not shown, for this surface has never been observed experimentally. Note that the Si(111)-$1\times1$ structure observed above the $7\times7\rightarrow1\times1$ phase transition temperature is not related to the static $1\times1$ structure reported here. It is actually caused by a thermally-average motion of atoms. The surface energy of the ideal (relaxed) $1\times1$ surface monotonically decreases from 127.1 to 102.0~meV/{\AA}$^2$ for Si(111), and from 93.5 to 77.6~meV/{\AA}$^2$ for Ge(111) when increasing the surface strain from $-0.04$ up to 0.04 (compressive surface stress). Surface energies of all other reconstructed surfaces show an opposite trend. They increase with strain (tensile surface stress) as seen in Fig.~\ref{fig3}. According to elasticity theory of solids, a non-vanishing derivative of the energy with respect to strain at $\epsilon=0$ indicates an intrinsic driving force for explosion or implosion of the crystal. This is not the case for surfaces since they couple to the underlying bulk, and therefore are not free to expand or contract infinitely in response to a surface stress.

%
\begin{table}
\caption{\label{tab1}Absolute surface energies and stresses (meV/{\AA}$^2$) for various Si(111) and Ge(111) surface reconstructions calculated using DZP basis at zero strain. Previously reported data are also shown for comparison (footnoted values).}
\begin{ruledtabular}
\begin{tabular}{lddddd}
                  &\multicolumn{2}{c}{Energy, $\gamma_\mathrm{rec}$}&  &\multicolumn{2}{c}{Stress, $\sigma_\mathrm{rec}$}\\
Reconstruction                         & \multicolumn{1}{c}{Si} & \multicolumn{1}{c}{Ge} &  & \multicolumn{1}{c}{Si} & \multicolumn{1}{c}{Ge} \\
\hline
\multirow{2}{*}{$1\times1$, unrelaxed} &  116.95                  &   86.45                  &  &                          &           \\
                                       &  113.57\footnotemark[1]  &   82.37\footnotemark[1]  &  &                          &           \\[1.0em]
\multirow{3}{*}{$1\times1$, relaxed}   &  112.33                  &   85.53                  &  &  -42.94                  &  -27.09   \\
                                       &  108.58\footnotemark[1]  &   81.12\footnotemark[1]  &  &  -39.31\footnotemark[2]  &           \\
                                       &  113.99\footnotemark[2]  &                          &  &                          &           \\[1.0em]
$3\times3$ DAS                         &   94.85                  &   70.45                  &  &  215.07                  &  148.33   \\[1.0em]
$5\times5$ DAS                         &   92.27                  &   68.83                  &  &  201.72                  &  141.28   \\[1.0em]
\multirow{2}{*}{$7\times7$ DAS}        &   92.42                  &   68.73                  &  &  191.80                  &  135.23   \\
                                       &   84.86\footnotemark[1]  &   63.65\footnotemark[1]  &  &                          &           \\[1.0em]
\multirow{2}{*}{$c(2\times8)$}         &   95.00                  &   68.62                  &  &                          &           \\
                                       &   87.98\footnotemark[1]  &   63.02\footnotemark[1]  &  &                          &           \\[1.0em]
\multirow{2}{*}{$2\times2$}            &   95.32                  &   68.79                  &  &  144.38                  &   95.65   \\
                                       &  105.34\footnotemark[2]  &                          &  &  141.51\footnotemark[2]  &           \\[1.0em]
$\sqrt{3}\times\sqrt{3}$               &   95.70                  &   71.09                  &  &  141.23                  &   98.40   \\
\end{tabular}
\end{ruledtabular}
\footnotetext[1]{From Ref.~\onlinecite{stekolnikov-prb-65-115318}.}
\footnotetext[2]{From Ref.~\onlinecite{vanderbilt-prl-59-1456}, assuming $a_\mathrm{Si}=5.420$~{\AA}.}
\end{table}

Looking at the surface energy plots for Si(111) in Fig.~\ref{fig3}(a), they can clearly be grouped into two sets. The first set includes $3\times3$, $5\times5$ and $7\times7$ DAS reconstructions, while the second set includes $c(2\times8)$, $2\times2$, and $\sqrt{3}\times\sqrt{3}$ structures. As explained below, such distinction stems from different building blocks that make their respective reconstructions. Analogous sets are observed in Fig.~\ref{fig3}(b) for Ge(111) as well. The only exception being that of the $\sqrt{3}\times\sqrt{3}$ surface energy data (adatom-based structure with the highest density of adatoms), which is somewhat less stable than other members of the second group.  Comparing $\gamma_\mathrm{rec}(\epsilon_\mathrm{b})$ curves from Figs.~\ref{fig3}(a) and \ref{fig3}(b), one readily conclude that the formation of simple adatom-based Ge(111) reconstructions compete with DAS-based counterparts near the region of zero strain, and that each structure type is prominent depending on the strain sign. On the other hand, Si(111) DAS-based structures are the most stable across most of the strain interval. This indicates that additional Si reconstruction elements like dimers and stacking faults are particularly important in the stabilization of the Si(111) surface.

In Table~\ref{tab1} we report absolute surface energies (DZP basis) and stresses (DZP-corrected SZ-DZP calculations) for various Si(111) and Ge(111) surface reconstructions at zero strain. Surface energy curves, $\gamma_\mathrm{rec}(\epsilon)$, were corrected to match DZP results at $\epsilon=0$. Surface stress values were taken from the first-order variation of $\gamma_\mathrm{rec}(\epsilon)$ at equilibrium,\cite{bechstedt-psp-2003}

\begin{equation}
\sigma_\mathrm{rec}=\left( \gamma_\mathrm{rec} + 
\frac{\partial\gamma_\mathrm{rec}}{\partial\epsilon_\mathrm{u}} \right)_{\epsilon=0},
\end{equation}
where $\epsilon_\mathrm{u}$ is the uniaxial strain (that relates to $\epsilon_\mathrm{b}$ through Eq.~\ref{eq:bi-uni}), and $\gamma_\mathrm{rec}(\epsilon_\mathrm{u})$ was approximated to a quadratic function that was fitted to the first-principles data.

Table~\ref{tab1} shows that unstrained Ge(111) with $c(2\times8)$ reconstruction is correctly predicted to be the ground state. Indeed, the experiments show that clean (111) surfaces of Ge samples have a $c(2\times8)$ structure.\cite{becker-prb-39-1633} 

According to our calculations, strain-free Si(111) surfaces with $5\times5$ and $7\times7$ DAS reconstructions are the most stable and they are almost degenerate. There is however a small preference for the $5\times5$ reconstruction by about 0.15~meV/{\AA}$^2$, which is just beyond the estimated error of 0.1~meV/{\AA}$^2$. The obtained ordering is at variance with the experiments. Clean Si(111) samples obtained after a high temperature annealing and a slow cooling stage down to room temperature, normally show a $7\times7$ DAS reconstruction. It should be noted though, that there is indirect experimental evidence that the surface energy of Si(111)-$5\times5$ is rather low and close to that of the $7\times7$ surface. Indeed, the $5\times5$ reconstruction has often been observed close to surface defects such as steps.\cite{kim-prb-81-085312} Further evidence was found during the phase transition from the metastable $2\times1$ reconstruction to $7\times7$ DAS, which proceeds through the formation of an intermediate $5\times5$ DAS surface structure.\cite{zhao-ss-418-132}

\begin{table}
\caption{\label{tab2}Absolute surface energies $\gamma_\mathrm{rec}$ at zero-strain (meV/{\AA}$^2$) for DAS-based reconstructions of Si(111) and Ge(111) surfaces calculated using the combined SZ-DZP basis.}
\begin{ruledtabular}
\begin{tabular}{ldd}
Reconstruction        & \multicolumn{1}{c}{Si} & \multicolumn{1}{c}{Ge} \\
\hline
$3\times3$ DAS        &   94.37   &   70.11   \\
$5\times5$ DAS        &   91.88   &   68.57   \\
$7\times7$ DAS        &   92.06   &   68.49   \\
$9\times9$ DAS        &   92.48   &   68.58   \\
\end{tabular}
\end{ruledtabular}
\end{table}

The energies of $3\times3$, $5\times5$ and $7\times7$ DAS reconstructed Si(111) surfaces were also calculated using the DFT method in Ref.~\onlinecite{stich-prl-68-1351}. The authors found that the $7\times7$ DAS reconstruction was the ground state, only 2~meV/{\AA}$^2$ and 3~meV/{\AA}$^2$ more stable than $5\times5$ DAS and $3\times3$ DAS surface reconstructions, respectively. However, these calculations were shown to suffer from poor BZ sampling, that once adequately corrected, resulted in a $5\times5$ DAS ground state 5~meV/{\AA}$^2$ more stable than the $7\times7$ DAS reconstruction.\cite{needels-prl-71-3612} Still, it is not clear why despite the $7\times7$ being the most commonly observed surface reconstruction in Si(111) after thermal treatment, the calculated Si(111)-$5\times5$ surface energy is actually predicted to be the ground state. Among the possible reasons for such disagreement one could have an insufficient number of layers considered in the slab, a poor treatment of the exchange and correlation energy, the neglect of entropy effects or the need for a better description of the electron density and potential.

To have an idea of the effect of the basis-choice to the surface energy, Table~\ref{tab2} lists absolute surface energies of various DAS reconstructions on Si(111) and Ge(111) calculated with SZ-DZP basis at zero strain. From a comparison with Tab.~\ref{tab1} we conclude that the energy difference between Si(111)-$7\times7$ and Si(111)-$5\times5$ reconstructions decreases by less than 0.5~meV/{\AA}$^2$ upon increasing the basis size. This result may direct us on future studies.

The magnitude and sign of the surface stress is directly related to the details of chemical bonding at the surface. It is possible to obtain a qualitative estimate of the impact of specific reconstruction elements to the overall surface stress. This is done by conceiving and inspecting \emph{test models} that include elemental structures of interest, even if the resulting surfaces are not observed experimentally. To this end the following test models were produced (i) $1\times1$ SF, $2\times2$ SF and $\sqrt{3}\times\sqrt{3}$ SF reconstructions, where stacking faults (SF) are introduced between the first and second surface layers, and (ii) $3\times3$, $5\times5$ and $7\times7$ dimers stacking fault (DS) structures, that basically are DAS counterparts after removing adatoms.

\begin{table}
\caption{\label{tab3}Calculated Si(111) surface stresses $\sigma$ (meV/{\AA}$^2$) for various experimentally observed reconstructions, compared to those of conceived model structures.}
\begin{ruledtabular}
\begin{tabular}{ldld}
 \multicolumn{2}{c}{Observed}            & \multicolumn{2}{c}{Test models}           \\
Reconstruction & \multicolumn{1}{c}{$\sigma$} & Reconstruction & \multicolumn{1}{c}{$\sigma$} \\
\hline
 $1\times1$               &    -42.94    & $1\times1$ SF               &    -1.06    \\
 $2\times2$               &    144.38    & $2\times2$ SF               &   163.85    \\
 $\sqrt{3}\times\sqrt{3}$ &    141.23    & $\sqrt{3}\times\sqrt{3}$ SF &   163.81    \\
 $3\times3$ DAS           &    215.07    & $3\times3$ DS               &   174.43    \\
 $5\times5$ DAS           &    201.72    & $5\times5$ DS               &   138.98    \\
 $7\times7$ DAS           &    191.80    & $7\times7$ DS               &   113.73    \\
\end{tabular}
\end{ruledtabular}
\end{table}

The comparison between surface stress calculations for observed and test model reconstructions is summarized in Tab.~\ref{tab3}. Several conclusions can be drawn from the data -- firstly, comparing the stresses of $1\times1$, $2\times2$ and $\sqrt{3}\times\sqrt{3}$ Si(111) surfaces with their counterpart values from SF structures, one may conclude that stacking faults produce a tensile contribution to the surface stress. Secondly, given that both DAS and DS structures have dimers located at the boundary of half unit cells, when their size is increased from $3\times3$ to $5\times5$ and then to $7\times7$ the relative contribution of dimers to the surface stress obviously decreases. This is accompanied by a decrease in the surface stress, meaning that the dimers in DAS reconstructions also produce a tensile contribution. Thirdly, when adatoms are added to the Si(111) truncated $1\times1$ surface to produce $2\times2$ and $\sqrt{3}\times\sqrt{3}$ reconstructions, they produce a strong tensile stress. Conversely, when adatoms are removed from surfaces having DAS reconstructions, the tensile strain decreases.

\begin{figure*}
\includegraphics[width=15.0cm]{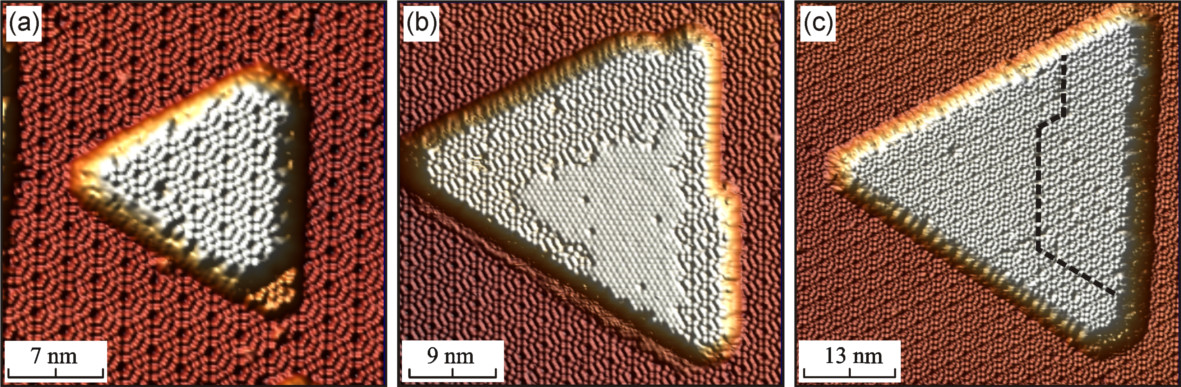}
\caption{\label{fig4}(Color online) STM images obtained during different formation stages of a germanium wetting layer on Si(111) at $T=400^\circ$C (quasi-equilibrium growth conditions). (a) Ge island formed during the initial stage exhibiting a $7\times7$ reconstruction with $\sqrt{S}/H=11.31$. At this stage, the Ge total coverage was 0.6~BL. (b) On rare occasions one can observe Ge islands with a $\sqrt{3}\times\sqrt{3}$ reconstructed top. Here $\sqrt{S}/H=24.33$ and the Ge total coverage was 0.4~BL. (c) During lateral growth, top surface reconstructions transform to $5\times5$. The dashed line highlights the domain boundary between $7\times7$ (left) and $5\times5$ (right) surfaces. The Ge total coverage was 0.8~BL and for this island $\sqrt{S}/H=31.68$. All islands in the figures are 3-BL thick.}
\end{figure*}

In summary, all reconstruction elements, \emph{i.e.} dimers, adatoms and stacking faults produce a tensile contribution to the surface stress. A tensile contribution to the surface stress is also expected from the reconstruction elements on (111) surface of germanium. The origin of compressive stress in the truncated Si(111) and Ge(111) with $1\times1$ periodicity lies at the Coulomb repulsion among the two-dimensional array of surface dangling bonds.\cite{bechstedt-psp-2003} We also note from Fig.~\ref{fig3} that the expected structure change is $c(2\times8)\rightarrow7\times7\rightarrow5\times5$ with increasing compressive strain for both Si(111) and Ge(111) surfaces. This is because compression relieves the intrinsic tensile stress so that reconstructions with bigger tensile stress are the most stabilized.

\subsection{Experimental results and comparison with theory}

We begin by looking at surface structure transformations that take place during the early stages of Ge/Si(111) growth. It is useful to define the parameter $\sqrt{S}/H$ to characterize the shape of growing Ge islands, where $S$ is the area of their top terraces ($\sqrt{S}$  is a characteristic lateral size of the terrace), and $H$ is the height of the island. Thus, $\sqrt{S}/H$ relates to the ability of the island to relax elastically, that is, higher values correspond to islands with a highly strained Ge/Si(111) upper layers, while lower values correspond to a more relaxed one.

The growth of Ge on Si(111) proceeds according to the SK growth mode, consisting in the formation of a few monolayers thick wetting layer during a first stage, followed by the growth of individual 3D islands as further deposition proceeds.\cite{bechstedt-psp-2003,motta-ss-406-254} The thickness of the wetting layer depends on the rate of Ge deposition and temperature.\cite{teys-jcg-311-3898} When growing at quasi-equilibrium conditions, as it was the case for this work (slow deposition rate $10^{-3}$-$10^{-2}$~BL/min and high temperature, \emph{i.e.} 300-500$^\circ$C), the formation of the Ge wetting layer takes place upon growth and coalescence of neighboring 3-BL thick Ge islands. If a higher deposition rate and/or a lower temperature are used, the formation of Ge islands with such thickness becomes kinetically limited, and two-bilayer or even single-bilayer thick islands may form.

Figures~\ref{fig4}(a)-(c) show STM images of typical 3-BL thick Ge islands grown at $T=400^\circ$C at various stages of the wetting layer formation. The STM image of a typical island formed at the very beginning of Ge/Si(111) growth is shown in Fig.~\ref{fig4}(a). The strained Ge islands formed at this stage exhibited the $7\times7$ reconstruction pattern of the underlying Si substrate surface. Rarely at this stage one could find Ge islands with a $\sqrt{3}\times\sqrt{3}$ reconstructed top terrace as it is depicted in Fig.~\ref{fig4}(b). Following further growth, the islands increase their lateral size (although maintaining their heights), and upon reaching some critical size, the $5\times5$ reconstruction pattern appears on top terraces [see Fig.~\ref{fig4}(c)]. Finally, the Ge islands merge together to form a continuous film (the so called the wetting layer), fully $5\times5$ reconstructed with no traces of $7\times7$.

\begin{figure*}
\includegraphics[width=15.0cm]{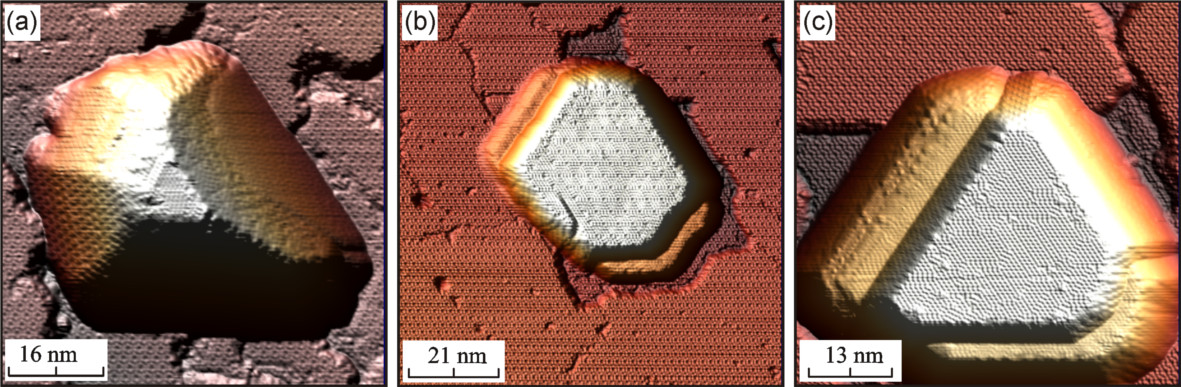}
\caption{\label{fig5}(Color online) STM images obtained during different stages of Ge growth on a Ge/Si(111) wetting layer at $T=400^\circ$C at quasi equilibrium conditions. This layer was invariably $5\times5$ reconstructed. (a) Ge islands formed during the early stages show small top terraces with $c(2\times8)$ and $2\times2$ reconstructions. The Ge total coverage here was 2.5~BL and for this particular island $\sqrt{S}/H=0.71$. (b), (c) Upon further growth, and depending on the $\sqrt{S}/H$ ratio, one observe both $7\times7$ and $c(2\times8)$ reconstructed terraces. The islands depicted have $\sqrt{S}/H=13.15$ and $\sqrt{S}/H=5.15$, respectively, and the Ge total coverage was 3.0~BL in both cases.}
\end{figure*}

Figures~\ref{fig5}(a)-(c) show STM images of typical 3D Ge islands grown at $T=400^\circ$C on top of the Ge wetting layer during various stages. Small 3D Ge islands formed early have well developed side facets and tiny top terraces.\cite{teys-jetpl-92-388} One of such 3D islands is shown in Fig.~\ref{fig5}(a), with its top terrace exhibiting $c(2\times8)$ and $2\times2$ domain reconstructions. Upon further growth, these 3D Ge islands increase in size, both laterally and vertically. Some of them exhibit a $7\times7$ reconstruction on their top surfaces [see Fig.~\ref{fig5}(b)], while others show a $c(2\times8)$ and $2\times2$ structures as shown in Fig.~\ref{fig5}(c). Note that the observation of $c(2\times8)$ reconstructed top layers indicates that they reached a relaxed state. This is the strain-free stable structure for clean (111) surfaces in pure Ge.\cite{becker-prb-39-1633}

The calculated surface phase diagram shown in Fig.~\ref{fig3}(b), accounts rather well for the observed surface structure transformation sequence that was just described. The Ge islands that are formed before completion of the wetting layer are compressively strained due to lattice mismatch (although some strain must be relieved at their edges). According to Fig.~\ref{fig3}(b) the effective built-in strain in such islands should be within $-0.01<\epsilon_\mathrm{b}<0.0$. In this strain range the $7\times7$ reconstruction is the most favorable state. As the islands increase in size, their ability to relax laterally decreases. Thus, wider Ge islands and particularly the continuous wetting layer, accumulate more compressive strain and that induces the $7\times7\rightarrow5\times5$ phase transition in agreement with Fig.~\ref{fig3}(b) for $-0.04 <\epsilon_\mathrm{b}<-0.01$.

When it comes to the formation of 3D Ge islands on top of the wetting layer, it is important to follow their $\sqrt{S}/H$ ratio. Initially, only islands with tiny top terraces and very small $\sqrt{S}/H$ values can be found. Due to their small lateral dimensions, these islands effectively release the compressive strain at their top layers. According to Fig.~\ref{fig3}(b), their $c(2\times8)$-reconstructed terraces indicate that they are mostly relaxed or eventually could accumulate some tensile strain. Larger 3D islands show either $c(2\times8)$ or $7\times7$ reconstruction, and again, we find a clear correspondence between $\sqrt{S}/H$ ratios and the observed terrace reconstructions: islands with lower $\sqrt{S}/H$ show relaxed $c(2\times8)$ terraces while islands with higher $\sqrt{S}/H$ exhibit compressive $7\times7$ terraces.

Our results indicate that the $2\times2$ reconstruction is metastable across the studied strain domain. Its formation during Ge/Si(111) growth, must be then driven by kinetics (not considered in this work). This structure was however frequently observed along with the $c(2\times8)$ reconstruction, and this is in line with the small calculated formation energy difference between these surfaces. The Ge(111)-$\sqrt{3}\times\sqrt{3}$ reconstruction with adatoms at $T_4$ positions is strongly unfavorable across the whole $-0.04$-$0.04$ strain range. The surface energy was therefore recalculated assuming $H_3$ high-symmetric positions for adatoms, although resulting in an even less stable structure. It is possible that the $\sqrt{3}\times\sqrt{3}$ structure is not made of Ge species only, and some Si intermixing could play an important role on its stabilization.

Another interesting system to observe the influence of strain on surface reconstructions is a stepped surface. The high-index regular stepped Si$(7\,7\,10)$ surface, ideally consists of (111) terraces comprising half of $7\times7$ unit cells in width and steps with height of three (111) interplanar spacings.\cite{teys-ss-600-4878} The regularity of this surface makes it an ideal substrate for growing nanodots and nanowires.\cite{zhachuk-ss-565-37} It should be noted that its orientation and atomic structure was the source of much controversy as it was formerly identified with a Si(557) surface.\cite{kirakosian-apl-79-1608,henzler-tsf-428-129,*zhachuk-prb-79-077401} Presently, the most consensual atomic model and the one that agrees better with the available experimental data is that reported in Ref.~\onlinecite{teys-ss-600-4878}.

\begin{figure}
\includegraphics[width=5.5cm]{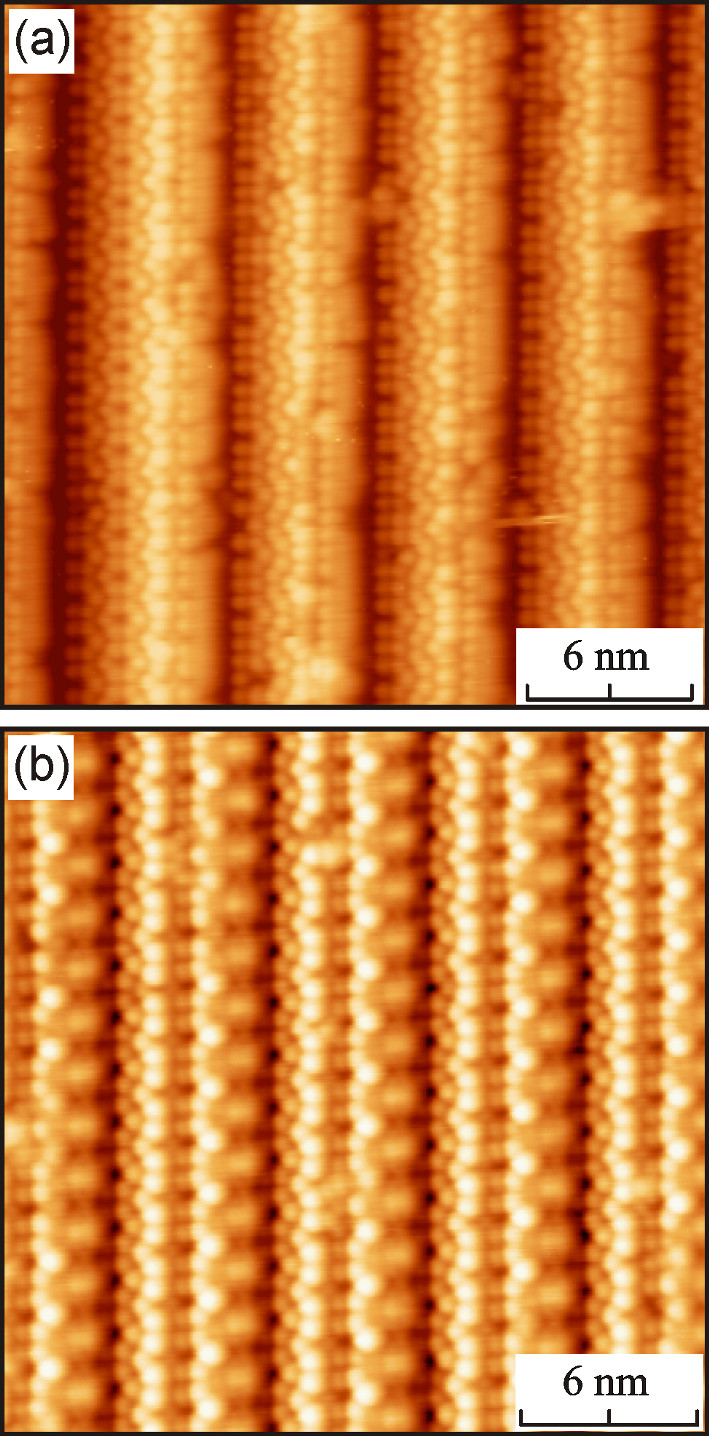}
\caption{\label{fig6}(Color online) STM images of a stepped Si$(7\,7\,10)$ surface with (111) terraces. (a) Clean Si$(7\,7\,10)$ surface with triple steps. All terraces show a $7\times7$ reconstruction except the one on the left, which has $5\times5$. (b) Si$(7\,7\,10)$ surface after adsorption of 0.3~BL Ge at $T=600^\circ$C. Triple steps are split into single and double steps. All terraces exhibit a $5\times5$ reconstruction.}
\end{figure}

Whereas Si(111) flat samples demonstrate surfaces almost fully $7\times7$-reconstructed, the (111) terraces of the Si$(7\,7\,10)$ stepped surface sometimes show $5\times5$ structure patches.\cite{kim-prb-81-085312} This is depicted in Fig.~\ref{fig6}(a). According to Fig.~\ref{fig3}(a), the observation of a $5\times5$ structure means that the surface layers at the (111) terraces suffer a compressive strain. Indeed, unlike the flat (111) samples, the (111) terraces on stepped surfaces have a nearby free border at the step edges. The Si(111)-$7\times7$ surface exhibits tensile stress (Tab.~\ref{tab1}), therefore in the presence of a free border, the surface area tends to shrink. This effect leads to a shift of the strain state as depicted in Fig.~\ref{fig3}(a) towards the compressive side (leftwards), making the $5\times5$ surface structure the most stable. This reasoning is, however, qualitative as it overlooks the structure of the step edges. These should differ for $7\times7$ and $5\times5$ reconstructed (111) terraces. The $5\times5$ reconstruction of a Ge layer deposited on Si$(7\,7\,10)$ can be explained based on a similar reasoning [see Fig.\ref{fig6}(b)].\cite{zhachuk-pss-51-202}

Finally we note that theory predicts that Si(111) surfaces with considerable tensile strain (above 0.03) should show a $c(2\times8)$ reconstruction [\emph{cf.} Fig. 3(a)]. Such strain can be produced when growing Si on Ge(111) surfaces and this should be verified in future experimental studies.

\section{Conclusions}\label{sec:conclusions}

Si(111) and Ge(111) surface formation energies were calculated as a function of an applied biaxial strain for a wide set of experimentally observed surface reconstructions. The calculations were carried out within the local density functional framework, by applying several bi-axial strains to vacuum-separated slabs with appropriate surface reconstructions. Using these data, the surface stresses of Si(111) and Ge(111) having different reconstructions, were calculated at the strain-free state. The results indicate that dimers, adatoms and stacking faults all produce a tensile contribution to the surface stress.

The formation energy of the Si(111) unstrained surface with a $9\times9$ DAS reconstruction was also calculated. It is shown that the energy of this surface is indeed higher than corresponding energies of surfaces having $5\times5$ and $7\times7$ DAS structures. However, the Si(111)-$7\times7$ surface is found to be slightly higher in energy than the Si(111)-$5\times5$, in apparent contradiction with the observations. This inconsistency calls for further work on this problem.

Strain phase diagrams for Si(111) and Ge(111) surfaces were constructed, showing a direct relation between the equilibrium surface structure and the applied strain. Comparison of the theoretical results with STM data shows that the constructed phase diagrams can explain the surface structure transformations observed during growth of strained Ge/Si(111), as well as the resulting structures found on the stepped Si$(7\,7\,10)$. Therefore, the observation of surface reconstructions may be used as a probe to the local strain, provided that we have a surface-strain phase diagram like the one presented in Figure~\ref{fig3}. Finally, theory predicts that Si(111) should adopt a $c(2\times8)$ reconstruction in the presence of a tensile strain approximately above 0.03.

\begin{acknowledgments}
We thank the Novosibirsk State University for providing us with computational resources. This work was supported by Russian Foundation for Basic Research (Project Num. 12-02-01128a). JC would like to thank for funding the NanoTP Cost Action Ref. MP0901 and to the Fundação para a Ciência e a Tecnologia, Portugal (FCT) under the grant PEst-C/CTM/LA0025/2011.
\end{acknowledgments}

\end{document}